# A First-order Phase Transition to Metallic Hydrogen

Mohamed Zaghoo, Ashkan Salamat, and Isaac F. Silvera
Lyman Laboratory of Physics, Harvard University, Cambridge MA02138

The insulator-metal transition in hydrogen is one of the most outstanding problems in condensed matter physics. The high-pressure metallic phase is now predicted to be liquid atomic from T=0 K to very high temperatures. We have conducted measurements of optical properties of hot dense hydrogen in the region of 1.1-1.7 Mbar and up to 2200 K. We observe a first-order phase transition accompanied by changes in transmittance and reflectance characteristic of a metal. The phase line of this transition has a negative slope in agreement with theories of the so-called plasma phase transition.

Theory and experiment have worked closely together for decades on the problem of metallic hydrogen. In 1935, 80 years ago, Wigner and Huntington (**WH**) predicted that solid molecular hydrogen would transform to atomic metallic hydrogen (**MH**) if pressurized to a quarter of a million bars (1 Mbar=100 GPa) [1]. MH is predicted to have spectacular properties such as room temperature superconductivity [2], possible metastability, and a prediction that the megabar pressure atomic metallic phase may be a liquid at T=0 K [3]. If metastable so that it exists at ambient conditions, MH would revolutionize rocketry as a remarkably light and powerful propellant [4]. In addition, the high-pressure molecular phase is also predicted to be metallic and superconducting [5,6]. Currently, static experiments to over 300 GPa have failed to reveal MH [7-11]. Modern calculations yield a transition pressure of 400-500 GPa for MH (see ref. [12,13]). After briefly reviewing recent and historical developments we present innovative new methods that allow measurements of transmittance and reflectance of hydrogen statically pressurized and heated to a metallic phase.

In recent decades a remarkable change has developed for the predicted phase diagram of hydrogen, shown in Fig. 1. The hydrogen melting line was predicted to have a non-classical dome shape with a maximum at ~ T=1000 K [14], confirmed by experiment [15-18]. The melting line was extrapolated to zero Kelvin to be consistent with earlier predictions (dashed line in the Fig. 1) [3]. At high pressures, above the melting line in the liquid molecular hydrogen field there is a predicted line of a first-order liquid-liquid phase transition to atomic metallic hydrogen, involving dissociation of the molecules [19], named the plasma phase transition or **PPT**. Early considerations by Landau and Zeldovich [20] discussed a first-order insulator-metal transition **(IMT)** in mercury with a critical point. This was studied for dense matter by Norman and Starostin [21] (theories for hydrogen will be discussed ahead). The PPT line has also been extrapolated to meet the melting line; for higher pressures, solid molecular hydrogen melts to liquid MH. The PPT is a Mott transition to a metal, whereas, for example, a low to modest density of hydrogen can electrically conduct as a plasma at very high temperatures [22], but is not a metal! A common IMT involves overlapping of energy bands; in disordered systems there are mobility bands. Mott has shown that metallization of disordered samples satisfy a minimum metallic conductivity condition [23]. As one traverses to ever-higher temperatures at high density there is no phase line that separates a low temperature metal from a high-temperature plasma [24]; a well-defined Fermi surface is a reasonable demarcation. Such regions are critical to understanding the state of most of the hydrogen in planetary systems, where it is estimated that 60-70% of the planetary mass in our solar system is dense fluid metallic hydrogen [25]. An important question is: does dense liquid hydrogen metallize as a continuous transition, or does it undergo a sharp first-order transition with a discontinuous increase in density? The importance of this question to our understanding of the structure of the astrophysical gas giants Saturn and Jupiter cannot be overstated [26] and is answered in this letter.

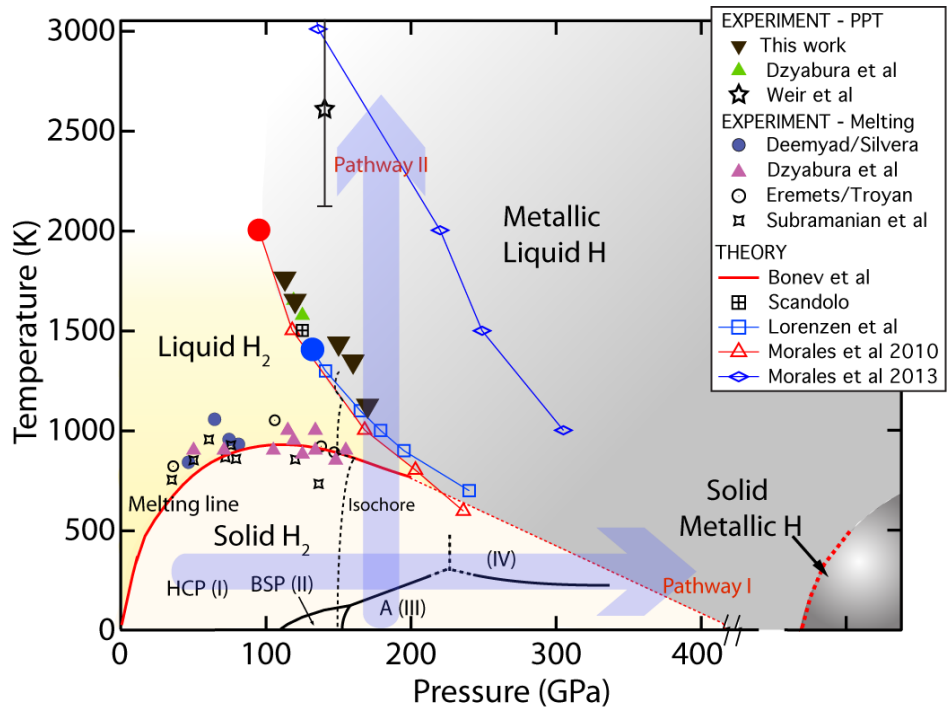

Fig. 1. Phase diagram showing several theoretical predictions of the plasma phase transition, as well as established phases in the solid at lower temperatures, and the theoretical melting line with confirming experimental data. Our pressure/temperature plateau data points (solid black triangles) delineate a phase boundary and optical measurements confirm the metallic nature of this transition. Circles at the end of the theoretical phase lines are critical points. An isochoric thermodynamic path is shown. The random uncertainty for plateau temperatures is $\pm 25\,K$ and falls within the size of the symbols; however there may be systematic uncertainties (see text).

The modern predictions in Fig. 1 show a broad field of liquid MH in the phase diagram, ranging from high temperature to zero Kelvin (gray region). In this figure two Pathways are indicated: I and II. Pathway I is at lower temperatures in the solid state; metallization requires static pressures not yet achieved on hydrogen in a diamond anvil cell (**DAC**). However, four phases have been identified, named I, II, III [27-29], and IV [9,10]. These were phase transitions of orientational order of the molecules in which the solid remained insulating. There have been several earlier reports of the observation of metallic hydrogen under static pressure conditions [9,30-33] but none have found acceptance in the scientific community [34-38].

The broad field of MH shown in Fig. 1 can also be entered at lower more accessible pressures along Pathway II, but very high temperatures are required, a challenge for high pressure hydrogen in a DAC. In this letter we describe our recently reported [39] achievement of producing MH at high P,T in a DAC.

Early theory, focused on planetary modeling, predicted a critical point for the PPT at ~15,000 K and ~60 GPa [40]. Density functional theory (**DFT**) [41-43] predicted somewhat higher pressures (**P**), and temperatures (**T**) almost an order of magnitude lower, shown in Fig. 1. Morales et al [44] then included nuclear effects and found an increase in the PPT pressure of ~100 GPa. A diffusion Monte Carlo calculation indicated that pressures ~ 650 GPa would be required [45], and subsequent such analysis predicts metallization a few hundred GPa lower [46] (not shown in Fig.1). More recently, DFT computational approaches have been benchmarked [47] with the result that the theoretical curves should fall in the original lower P/T regions shown in Fig. 1. An experimental determination of the PPT is still lacking and would resolve these large differences in theoretical predictions.

Hydrogen and deuterium have also been studied by dynamic shock wave techniques that achieve high pressures at much higher temperatures (~10-50x$10^3$ K) for periods of 10-100 nanoseconds [48-52]. Optical measurements show a continuous rising reflectivity consistent with metallic behavior. A phase transitions such as the PPT has not been observed, likely because their thermodynamic pathway precludes this. Such measurements were unable to distinguish between atomic or molecular conductivity. Experiments using reverberating shock waves by Weir, Mitchell, and Nellis [53] found the conductivity of hydrogen to saturate at a pressure of 140 GPa and a calculated temperature estimated to be 2500 to 3500 K, with minimum metallic conductivity [54]. Shock experiments with optical measurements find a much lower density for the transition to metallic behavior, possibly due to higher temperatures. Fortov et al [55] observed a density change in shocked deuterium, believed to be due to the PPT (not shown in Fig. 1, the phase diagram of hydrogen). Neither this, nor the experiment of Weir et al measured temperature, and the density data in Fortov et al's work is too sparse to conclude a discontinuous behavior.

Our optical determination of metallization of hydrogen is based on a Drude model. In this model a metal has a plasma frequency. For frequencies higher than the plasma frequency the metal is transparent, while for lower frequencies the transmittance decreases and the reflectance approaches 100%. We cannot directly measure the plasma frequency because the diamond anvils of the DAC are opaque in the region where the plasma frequency is expected (around 20 eV) [56]). We have studied optical properties using several laser lines (CW laser light at 514 nm, 633 nm, 808 nm, and 980 nm), almost a factor of two in wavelength using the optical system shown in the Supplementary Information, Fig. SI1 The optical method is often used in dynamic [48,51] and static high pressure studies of IMTs [57-59]. Measurement of electrical conductivity as an experimental method for a first observation of MH was rejected as the metallic leads might contaminate the hydrogen [38].

A DAC is an adjustable isochoric (fixed volume) cell. For a given volume, as temperature is raised, the sample follows an almost vertical thermodynamic path shown in Fig. 1. Heating along this trajectory, hydrogen first melts to a molecular liquid, then at higher temperatures the PPT line is crossed and molecules are predicted to dissociate to liquid atomic MH. In a given phase, pressure increases weakly with temperature. However, since the slopes of the phase lines traversed are negative, the pressure falls back when crossing such lines, somewhat compensating the thermal pressure. We measure the pressure at room temperature as representative of the pressure at high temperature. Pressure is determined either by ruby fluorescence or the shift of the Raman active vibron of the hydrogen (see SI2).

The challenge of studying high P,T hydrogen arises from hydrogen diffusion into the diamond anvils, which then fail. At high P/T, hydrogen is very reactive and can diffuse into the metallic gasket or diamond; diamond anvils can embrittle and fail. We have developed methods to inhibit diffusion of hydrogen at high temperatures using pulsed laser heating and coating the diamonds with alumina as a diffusion barrier. Laser pulses are ~280 ns long; this is sufficient time to achieve local thermal equilibrium in the hydrogen (electron/ phonon relaxation times are of order 10 ps or less [60]), but too short for serious diffusion to ensue. We use a small table-top laser with pulse energies less than 2 mJ to heat the already dense hydrogen. We measure P, T and optical properties that confirm the long-predicted transition to a liquid metallic phase.

Hydrogen is pressurized in a DAC at room temperature. A rhenium gasket confines the sample, which in our many runs ranged from 10-30 microns in diameter. Since hydrogen is transparent one can laser heat an embedded absorber that in turn heats the hydrogen pressed against its surface, as was done in refs. [15,18]. The surface of the hydrogen sample was heated to peak temperatures as high as 2200 K. At high pressures the molecular hydrogen is a few microns thick, while we estimate that the heated part of the sample that metallizes is a few to tens of nanometers thick. The energy/pulse is small, so that the sample and diamonds remain at room temperature, on average. The temperature of the absorber and adjacent hydrogen is determined by collecting the blackbody (BB) irradiance and fitting to an appropriate curve based on the Planck radiation law to yield the peak temperature due to the heating pulse [61].

In earlier experiments by Dzyabura, Zaghoo, and Silvera [18] a laser heated absorber embedded in the hydrogen heated the hydrogen pressed against its surface. In this configuration a phase transition line was observed in agreement with some of the theoretical lines for the PPT. They measured <u>heating curves</u>, i.e. plots of peak temperature vs. heating power. For such curves, if there is no phase change the temperature increases monotonically with laser power. They observed plateaus at P,T values (constant T as laser power is increased) interpreted as being due to heat of transformation (energy goes into latent heat) or due to increases in reflectance, as both of these mechanisms require increased laser power to maintain the temperature. However, they presented no evidence of metallic behavior. In this letter our objective is to show that the $P_C/T_C$ values of the plateaus ($P_C$ and $T_C$ are the values at the plateaus; see curves in Fig. 2 and Fig. SI5) represent the PPT phase line, and that for $T \geq T_c$ the optical properties are that of a metal, i.e. MH. These P/T points are plotted in Fig. 1.

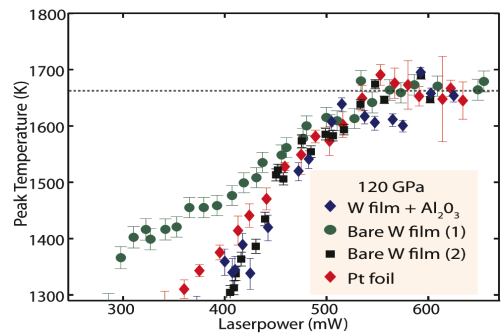

Fig. 2. Overlapping of plateaus on heating curves for several different laser absorber surfaces at a fixed pressure of 120 GPa. Each point represents the peak temperature achieved for a given laser power.

It is important to demonstrate that the plateaus arise from hydrogen and not the absorber. In Fig. 2 we show plateaus for absorbers made of platinum, tungsten (**W**), and W with a thin alumina layer to inhibit hydrogen diffusion and chemical activity (such as dissociation of molecular hydrogen on a metallic surface). At high P,T, tungsten forms tungsten hydride [62], whereas the protected W should be tungsten. There is excellent overlap in the data, establishing that the plateaus arise from hydrogen. Einaga et al [63] have demonstrated plateau-free heating curves with an inert medium on a gold absorber.

Here we report on a new method to measure the optical properties of transparent dense matter. We use a thin (~8.5 nm) semi-transparent film of tungsten (**W**) deposited on the diamond culet as our laser coupler; this enables transmittance and reflectance of the hydrogen to be measured. The diamond is insulated from the heated W by a 50 nm layer of (amorphous) alumina (Fig. SI3). Thus, one can <u>continuously</u> shine optical light on the sample while repetitively (20 kHz) heating the hydrogen with the pulsed laser. During the pulse the heated sample traverses the isochoric line shown in Fig. 1, rising and falling in temperature. With sufficient laser power the PPT region can be reached. The transmitted (or reflected) light is detected with a Si photo-detector with a 3 ns rise-time. The optical light transmittance (**Tr**) through the film and hydrogen, as well as the reflectance (**R**), is measured during the pulse. The waveform is recorded and averaged on an oscilloscope for several microseconds, synced to the laser pulse (the optical setup is shown in Fig. SI1). The pulsed laser light used for heating is filtered out. The Tr/R due to the hydrogen can be separated from that of the W film. The transmittance of the cell in the light path is temperature dependent (thermo-transmittance) and this is taken into account (see SI).

Fig. 3 shows the Tr/R beyond the plateau region. Data analysis requires a detailed description. First, because different laser wavelengths have different powers, we scale the Tr/R signals to 1 for each laser wavelength (for times before or after the laser pulse when the sample is at room temperature). This represents the transmittance and reflectance (we plot R+1) of the W, molecular hydrogen, and diamonds, so that losses due to the cell are normalized out. To get absolute Tr of the metallic hydrogen, we measure the transmitted signal for laser power pulses at a temperature $T_c$ on or above the plateaus (see Fig. 2) and these time traces are divided by the signals for temperatures less than $T_c$ just below the plateaus (for raw data see SI4). This is important, to remove the temperature dependence of the transmittance of the cell. With this normalization the transmission of the molecular hydrogen before it metallizes is 1 (see SI for detailed analysis). The reflectance is more complicated. We define $R_s$ as the ratio of the detected signal S(t) above $T_c$ to that below $T_c$, reflected off of the W film. For a thick film of MH this yields R. However, for a thin film the signal S contains light reflected off of the MH, plus light transmitted through the MH and reflected off of the tungsten, back through the film and to the detector, twice attenuated by the MH. We analyze the multifilm reflectance using Fresnel equations to extract R (see SI). The W layer may be converted to tungsten hydride (due to diffusion) [62] and its complex index of refraction is unknown. As a consequence we measured the reflectance of the tungsten layer at high pressure, which is needed for the analysis (see SI).

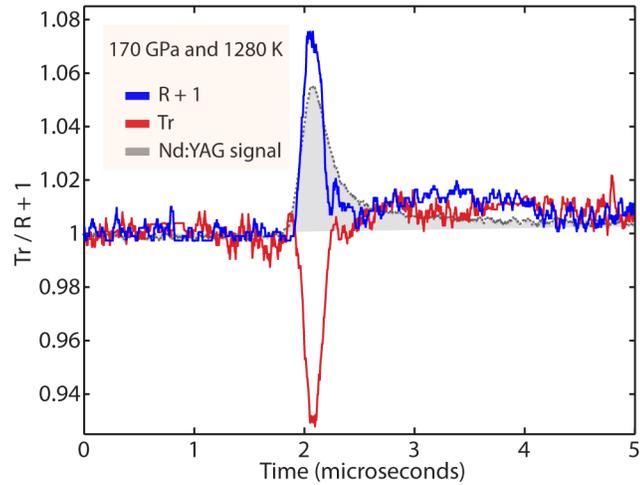

Fig. 3. The transmittance and reflectance (980 nm light) of hydrogen vs. time at 170 GPa and 1280 K when hydrogen is heated into the region beyond $T_c$. The trace of the 280 ns wide Nd:YAG pulse indicates the period of time when the hydrogen is hottest. The laser signal is on an arbitrary scale.

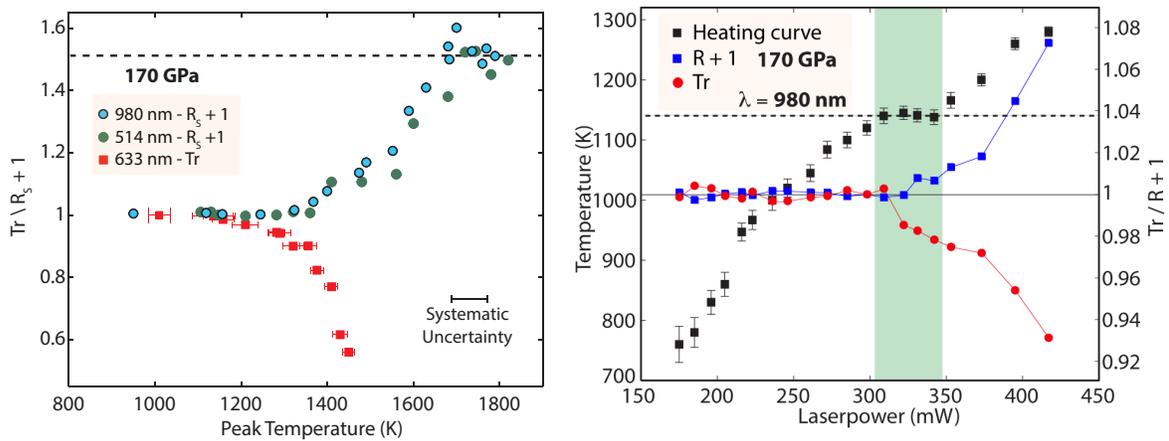

Fig. 4. Left: Transmission and normalized reflectance signal as function of temperature traversing from the thin MH film to the thick film limit and saturating the reflectance. Right: The heating curve at 170 GPa showing a distinct plateau at $1140 \pm 40$ K. Each point represents the peak temperature achieved for a given pulse power. The rising reflectance and falling transmittance are shown for powers corresponding to temperatures at the plateau (indicated by the vertical band) and higher. Peak temperature and laser power were not measured simultaneously for the left figure, giving rise to the indicated systematic uncertainty.

In total for this letter 11 runs were carried out (see SI), all consistent with each other. When the absorber film was overly heated it could deteriorate so that we conservatively studied the transition to MH mainly in the vicinity of the plateau where the MH film is thin, of order a few nm, and the changes in T/R values are only several percent (see scale in Fig. 3). However, at higher pressures, the plateau is at lower temperature and we safely heated substantially above the plateau temperature, Fig. 4, left. Since transmittance decreases exponentially with the absorption coefficient times the thickness (see SI) the explanation of decreasing transmission is that the inhomogeneous film of MH becomes thicker with increasing power. We see that as the film thickens (higher peak temperature) the reflectance is saturated to bulk values of about 0.55; this is consistent with shock measurements [49] on deuterium. From calculations of Tr/R for thin films using a dielectric function for hydrogen, the estimated thickness at saturation is ~30-50 nm. In Fig. 4, right, we show Tr/R+1 at different points on the heating curve in the thin film limit, demonstrating their variation as the plateau region is traversed. The drop in Tr and increase in R grows with temperature (or pulsed laser power). Note that the reflectance rises very slowly on the plateau, less than 0.01, compared to changes in the transmittance. Thus, we attribute the origin of the plateau to latent heat of transformation (rather than increased reflectance). Latent heat (due to dissociation) is a characteristic of a first-order phase transition that has been predicted for the PPT. Tr/R data for all of our studied wavelengths spanning a factor of ~2 in frequency (see Fig. SI6) are supportive of the Drude model of a metal.

We observed no evidence of interband transitions. The fact that the thin hydrogen film is inhomogeneous (i.e. not a slab) is considered in a theoretical paper along with estimates of the dielectric and Drude parameters by fitting Tr/R data [64]. Assuming full dissociation at 170 GPa and 1250 K, the DC Drude conductivity is found to be $(2.1\pm1.3)\times10^3(\Omega cm)^{-1}$, satisfying Mott's criterion for the minimum metallic conductivity for hydrogen.

We estimate that $\sim 1\times 10^{-5}$ mJ of energy is required to dissociate a monolayer with the area of our sample at 170 GPa or about $\sim 10^{-3}$ mJ for a 10 nm film (see SI for details). The energy in a laser pulse falling on the sample can easily be varied in small increments, so that it is well matched to heat the sample and scan a heating curve involving a phase change.

In summary, the results we report provide strong evidence for the IMT in dense liquid hydrogen. The PPT that has long been predicted by theory, but heretofore never observed. A phase line with a negative slope has been observed that separates the insulating phase from the higher temperature metallic phase. The changes in Tr and R at this transition, over a factor of almost two in wavelength, are supportive of a Drude model of a metal. Plateaus are associated with latent heat so that the line represents a first-order phase transition with coexisting phases that thicken as the temperature rises. A characteristic of the PPT is a critical point and discontinuity in density; our pressure range may have been too high to observe the critical point. Discontinuous changes in density characteristic of a first-order phase transition are not observable in an isochoric (fixed volume, thus fixed density) cell such as a DAC chamber. For the first time our results show that the transition to a conducting phase is not continuous as has been reported in shock experiments. At this point in our studies we cannot experimentally distinguish between a molecular metal and an atomic metal, but the correlation between our data and theory indicates atomic MH [41-43]. Temperatures in this set of experiments were too high to expect to observe metastable MH. We also do not expect to detect high Tc superconductivity at these temperatures. Future measurements at higher pressures, implying lower transition temperatures, may enable observation of these predicted properties.

We thank Vasya Dzyabura for earlier research on this subject and many discussions with Jacques Tempere and Nick van den Broeck, as well as Bill Nellis, and Takashi Yagi. We are grateful to Steve Byrnes who provided assistance with the multilayer film analysis. The NSF, grant DMR-1308641, the DOE Stockpile Stewardship Academic Alliance Program, grant DE-FG52-10NA29656, and NASA Earth and Space Science Fellowship Program, Award NNX14AP17H supported this research. Preparation of diamond surfaces was performed in part at the Center for Nanoscale Systems (CNS), a member of the National Nanotechnology Infrastructure Network (NNIN), which is supported by the National Science Foundation under NSF award no. ECS-0335765. CNS is part of Harvard University.

# A First-order Phase Transition to Metallic Hydrogen
## Supplementary Information

Mohamed Zaghoo, Ashkan Salamat, and Isaac F. Silvera
Lyman Laboratory of Physics, Harvard University, Cambridge MA02138

**DAC and Optical System**

The data on the plateaus and optical properties were obtained in 13 runs; details are given in Table I. The DACs are of the design described in Ref. (*1*) but made of Vascomax. In all runs, diamonds were type Ia, selected for low fluorescence with 75 or 100 μm culet flats. Samples were cryogenically loaded and then warmed to room temperature and removed from the cryostat to be mounted on an optical table for measurements. The layout of the optical system is shown in Fig. SI1. This system allows for pulsed laser heating with temperatures determined from the thermal irradiance of the pulse heated laser absorbers. This light is collected with a Schwarzschild objective, magnified and imaged on a 200 μm diameter spatial filter and then reimaged on the slits of a prism spectrometer, detected with an InGaAs diode array responsive in the 0.8 to 1.7 μm spectral region.

Thermal irradiances were fitted to a grey-body curve in the spectral region of 1.33-1.7 μm using techniques described elsewhere (*2*). The transfer function used to correct the spectral irradiance from our samples to the response of the optical system was measured using an ohmically heated thin tungsten film deposited on a diamond anvil in an inert environment, to best match the desired condition of the experiment. The high emissivity of the W films, ~4 times that of bulk foils resulted in higher signals and allowed for temperature measurements as low as 800 K, with a duty cycle of $\sim 1.5 \times 10^{-3}$. The system included an optical spectrometer used for pressure determination from ruby fluorescence or Raman spectra of the hydrogen vibron, shown in Fig. SI2. Above 1 Mbar, all the pressure points reported within this article are based on the hydrogen Raman shift as a pressure marker using a calibration from Ref. (*3*). Four different CW probe laser beams could be sequentially imaged on the sample by flipping of mirrors. Each probe laser was aligned to be collinear with the pulsed laser beam. In general the DAC was oriented so that the beams go through the hydrogen sample before hitting the semi-transparent W film (see Fig SI3). The transmittance and reflectance signals of these laser beams was detected by fast Si detectors; the output was recorded and averaged on a 100 MHz bandwidth multichannel oscilloscope, synced to the heating laser pulse. For reflectance measurements, the DAC was rotated 10 degrees from the normal so that the central mirror of the Schwarzschild did not block the reflected beam. Various important filters and dichroic mirrors are shown in Fig. SI1. A computer controlled beam stop and polarization rotator aided automated data acquisition. Fig SI3 shows the heart of the DAC with the multilayer films. For clarity this figure is not to scale.

| Runs | Pressures studied | Absorber material | Data collected |
|---|---|---|---|
| Run1 | 140-155 GPa | Pt foil+ Alumina | Plateaus, Ref. (2) |
| Run2 | 119-129 GPa | Pt foil | Plateaus, Ref. (2) |
| Run3 | 110-125 GPa | W thin film | Plateaus |
| Run4 | 120-130 GPa | W thin film | Testing /calibration |
| Run5 | 120-176 GPa | W thin film | Plateaus +testing optical data acquisition/analysis |
| Run6 | 120-125 GPa | W thin film | Plateaus |
| Run7 | 120-135GPa | W thin film | Plateaus + optical data |
| Run8 | 130-135 GPa | W thin film + Alumina | Plateaus + optical data |
| Run9 | 96 GPa | W thin film + Alumina | No optical data |
| Run10 | 106 GPa | W thin film + Alumina | No optical data |
| Run11 | 117 GPa | W thin film + Alumina | Diamonds failed |
| Run12 | 130 GPa | W thin film + Alumina | Plateaus and optical data |
| Run13 | 130-170 GPa | W thin film + Alumina | Plateaus and optical data |

Table I. A compilation of experimental runs that were carried out for measurements on high-pressure pulse heated hydrogen.

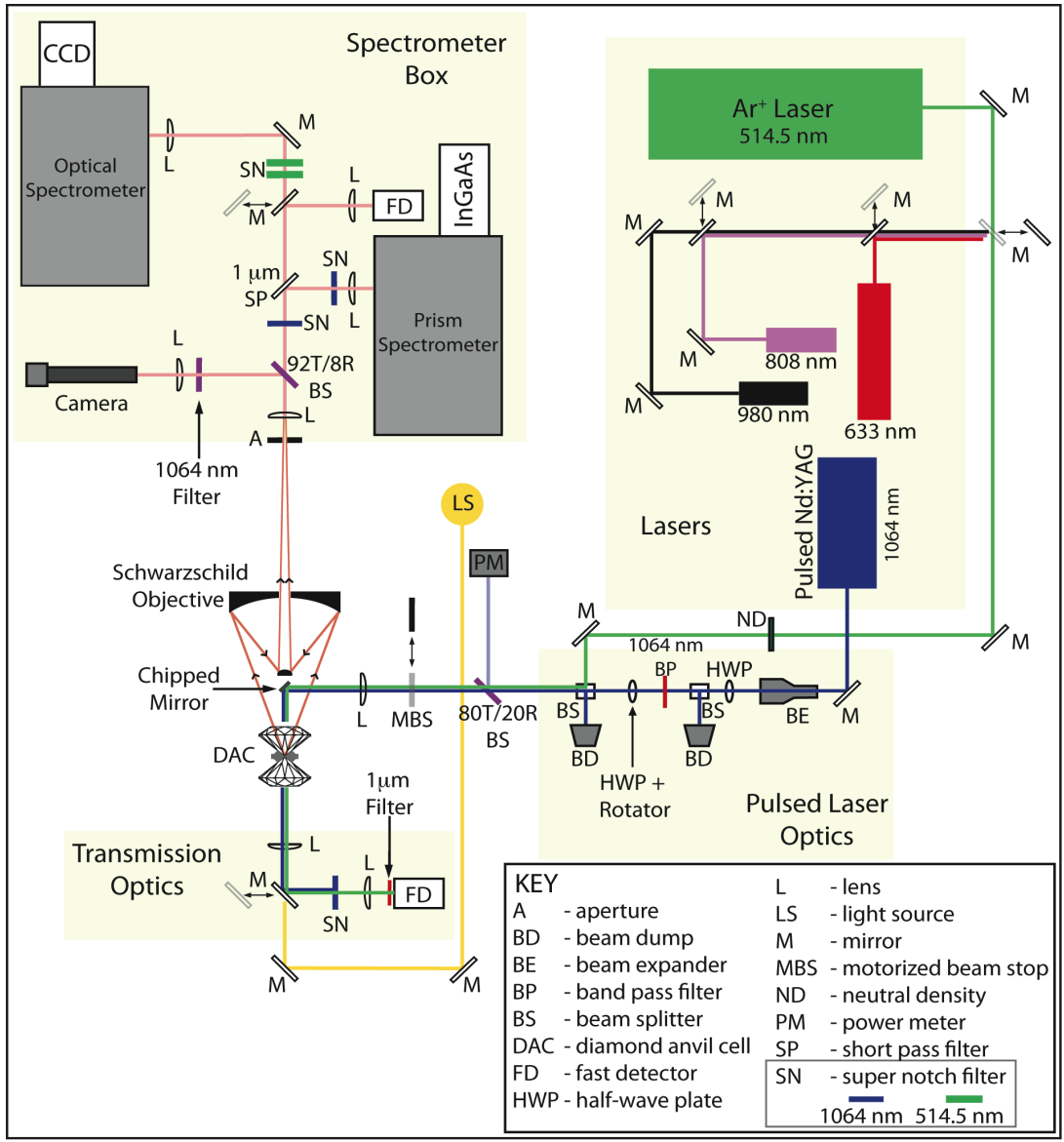

Fig. SI1. The optical layout for studies of pulse heated hydrogen.

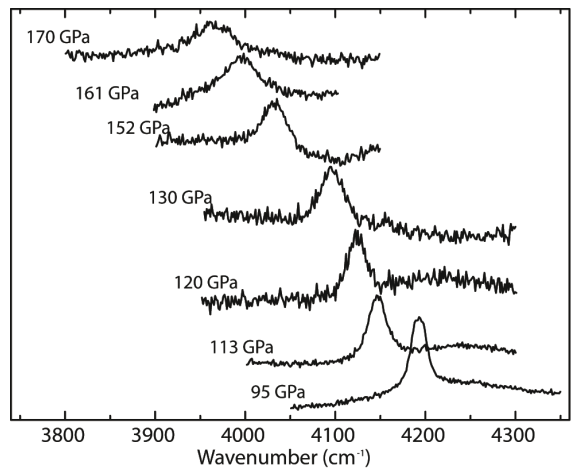

Fig. SI2. Raman spectra of the hydrogen vibron mode used for pressure measurements. Pressure was measured before and after each optical measurement.

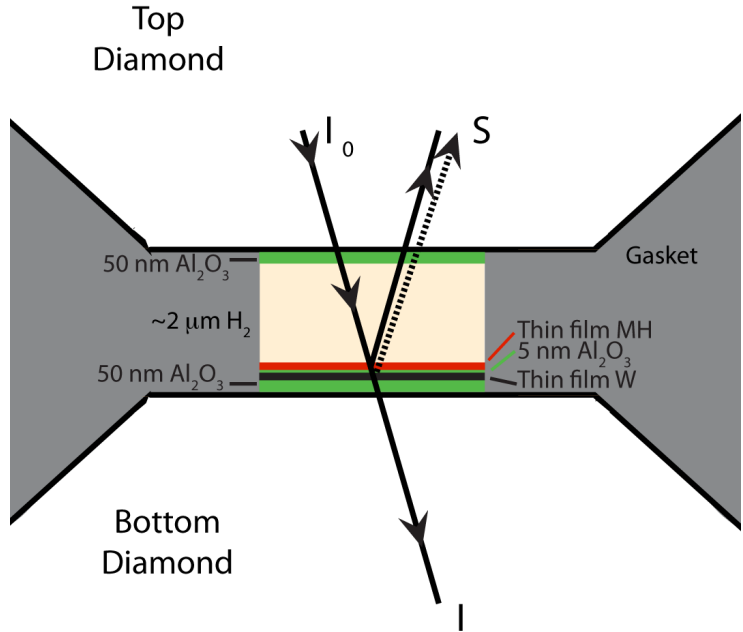

Fig. SI3. Interior of the cell showing the various layers and light vectors, including secondary reflections. The tungsten film is ~8 nm thick (layers are not to scale). For some samples a 5 nm layer of alumina was deposited on the W film. For clarity we do not show the bending of rays at interfaces.

**Transmittance and Reflectance Analysis**

In the DAC sample in Fig. SI3, only tungsten and MH have the dominant emissivity in the spectral region of interest, 0.8-1.7 µm, whereas that of diamond, amorphous alumina, and molecular hydrogen are negligible. Thus, the response of transmittance and reflectance signals are dependent mainly on the tungsten thermo-optical properties at lower temperatures. This is valid unless hydrogen undergoes a phase transition that changes its electronic/optical properties. The W film is much thinner than the optical attenuation length so it is uniformly heated and thermalizes in the order of ten picoseconds; its temperature essentially follows the power of 280 ns wide laser pulse. Moreover, the W film and alumina layers are much thinner than the wavelengths of the laser light (>500 nm), so that there is little phase shift of the light and no interference fringes in the layers. The layer of hydrogen adjacent to the W film is in local thermal equilibrium with the W during this relatively long pulse, with a small temperature offset when the absorber has a thin alumina coating. When the hydrogen is sufficiently hot to become metallic, we observe a dip in the transmittance, with an increase in the reflectance for higher temperatures (see Figs. 3,4). The raw data for such a situation is shown Fig. SI4a. The pulse energy flows mainly into the more massive molecular hydrogen (a few microns thick; a smaller portion flows into the diamond with the W coating). The molecular hydrogen has a large heat capacity and longer thermal time constant. Thus, the temperature of the molecular hydrogen rises and falls; after the laser pulse ends the temperature of the W follows that of the molecular hydrogen (consider injecting a finite amount of energy into a mass in a delta function in time; the temperature of the mass rises and falls with its thermal time constant). In Fig. SI4 we observe a thermal response of a few microseconds, much longer than the time of the heating pulse.

In order to isolate the optical properties of the hydrogen film adjacent to the tungsten, for a multilayer system such as depicted in Fig. SI3, one needs to calculate the transmitted and reflected electric fields using the Fresnel equations and squaring to get the transmittance and reflectance. We have done this for our optical pathway, simulating a metallic hydrogen film. We find that for nanometer thick films we can use an "intensity approximation", that is, propagate the intensity through the multifilm system. The approximation is rather good for Tr, but it overestimates the reflectance by ~10-15%; the thicker the MH film, the more accurate the approximation. We present the analysis in the intensity approximation here. Assume that the laser intensity incident on the DAC is $I_0$ and the detected power in transmittance is $I$ (see Fig. SI3), then

$$I(t,T) = I_0 Rf \exp[-(\bar{\alpha}_{diam} + \bar{\alpha}_{H_2} + \bar{\alpha}_{Alumina} + \bar{\alpha}_{W(t,T)} + \delta_{x(t)H}\bar{\alpha}_{H(t,T)})] \quad (1.1)$$

where we incorporate the reflective losses (due to the diamond tables) in $Rf$; $\bar{\alpha}_i = \alpha_i d_i$ with $d_i$ the thickness of layer i, $\delta_{xH}=1$ if $x=H$ and 0 if $x=H_2$, and t, T are time and temperature. We incorporate the layer thickness in $\bar{\alpha}_{H(t,T)}$, not only for ease of notation, but because we do not measure the thickness of the H film. Now consider a transmittance signal in time for a temperature T below the PPT transition temperature $T_c$, and call this $I(t,T<T_c)$ with $x=H_2$, and another for a temperature above the transition temperature $I(t,T>T_c)$ with $x=H$. Then, the transmittance at or above $T_c$ is

$$Tr = I(t,T>T_c)/I(t,T<T_c) = \exp[-\bar{\alpha}_{H(t,T>T_c)})] \quad (1.2)$$

For $T<T_c$, hydrogen is in the insulating molecular phase with $x=H_2$ and the transmittance is 1 (we have normalized out the transmission of the W film). A similar approach is used for the reflectance signals to measure the absolute reflectance of a layer of metallic hydrogen on top of a reflective W surface. In this case let $R_W$ be the reflectance off of the tungsten film (geometry of Fig. SI3) and R the reflectance off of the metallic hydrogen. Then the reflectance signals are

$$S(t,T>T_c) = I_0 Rf(R + Tr^2 R_W) \quad (1.3a)$$
$$S(t,T<T_c) = I_0 Rf R_W \quad (1.3b)$$

Normalizing gives

$$R_S = \frac{R + (Tr)^2 R_W}{R_W} \quad \text{or} \quad R = R_W[R_S - (Tr)^2] \quad . \quad (1.4)$$

As explained in the text we normalize $R_W$ to 1, so in this case eq. 1.4 becomes

$$R = [R_S - (T_R)^2]. \quad (1.5)$$

The second term in Eq. 1.3a or 1.4 for $T>T_C$ is due to the light that passes through the H film and is reflected back by the W film. When hydrogen is in the insulating phase $R$ is zero and $Tr=1$; with this normalization $R_S = 1$ (Eq. 1.4). In the thick film limit the transmission goes to zero and R=R$_S$. We plot reflectance of our MH films as $R+1$ for convenience.

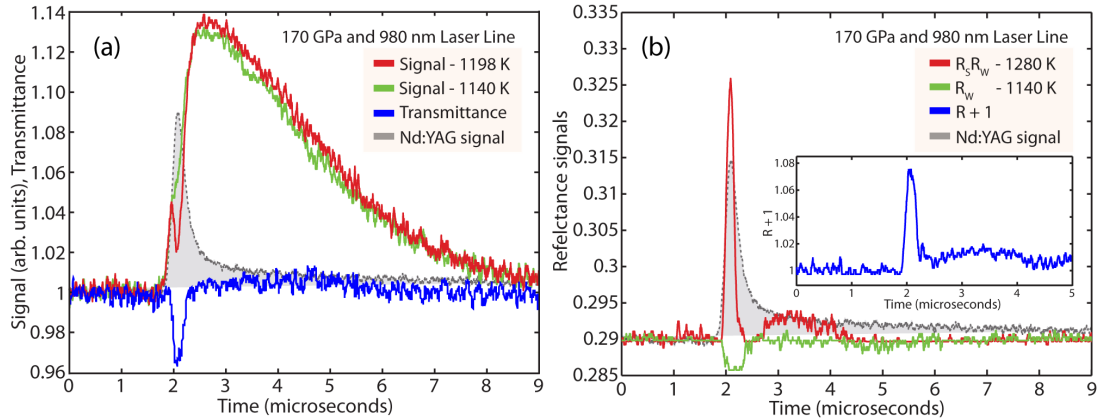

Fig. SI4. (a) Raw transmittance signals during the heating pulse at two different temperatures lower than and in the $T_c$ region. Normalized transmission of the hydrogen (blue trace) is the ratio of the high temperature to the low temperature signals. (b) Plots of $S(t,T>T_c)$ (red) and $S(t,<T_c)$ (green), both normalized to $I_0 Rf$. The inset shows R+1, where $R_S$ has been corrected for the secondary reflection off of the W film.

In order to determine the absolute value of R using Eq. 1.4, we need to determine $R_W$. We measured the transmittance through the tungsten film in the DAC at 170 GPa, and prepared another DAC without the W film but with diamonds coated with alumina and a gasket with the same size hole. Normalization gives the transmittance of W (we correct for the index of air in the gap between diamonds in the second DAC to have that of high pressure hydrogen). Normalizing we have $Tr_W = I(t,T<T_c)/I_0 Rf$ which enables the determination of $I_0 Rf$, using the measured signal. Then $R_W = S(t,T<T_c)/I_0 Rf$, and we find $R_W = 0.29$. This enabled us to extract the reflectance of MH from the measured reflectance of the multilayer sample.

The method of normalization (for example, Eq. 1.2) corrects for the temperature dependence of transmittance and reflectance of the heated W region. An example of raw data is shown in Fig. SI4.

Plateaus in the heating curves, attributed to the latent heat of dissociation of the PPT, are shown in Fig. SI5. The P,T points for the plateaus are plotted in Fig. 1.

Figure SI6 shows some examples of transmittance at different pressures. Due to experimental constraints the transmittance and reflectance measurements as a function of wavelength were not all conducted simultaneously, but sequentially. In some cases we measured the temperature as function of pulsed laser power, and subsequently used the power to imply the temperature. For some of the transmittance measurements we could not measure the temperature

simultaneously with the transmittance signal. This was because the longer wavelength laser light could not be completely filtered out of the blackbody light used to determine the temperature and the two contributions could not be separated. We have conducted systematic studies on the uncertainty of the temperature measured for a given laser power over a period of time. For time intervals of ~1 hour, for the same measured laser power, the temperature uncertainties are on the order of ±50 K. This systematic uncertainty is attributed to mechanical instabilities, where slight spatial shifts could change the position of the focus of the heating laser relative to the sample by few microns.

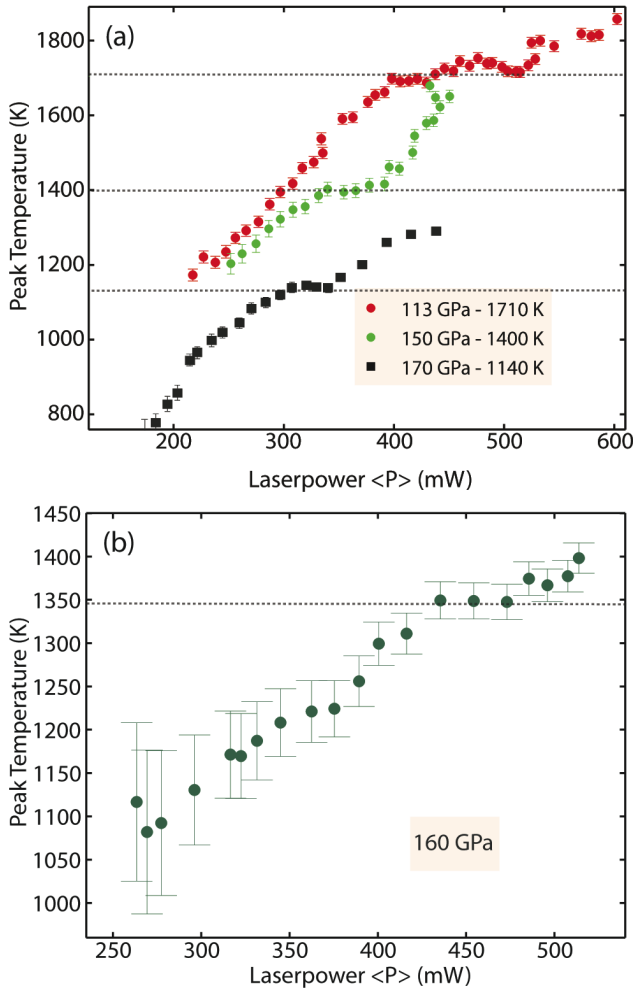

Fig. SI5. Several heating curves showing the rise in temperature of the hydrogen as power is increased for the different pressure points reported in Fig. 1. For every pressure reported, three different heating curves were measured, often in different runs (details in Table 1). The temperature uncertainty bars represent the standard deviation of the fitted temperatures; the straight lines at plateaus are guides to the eye.

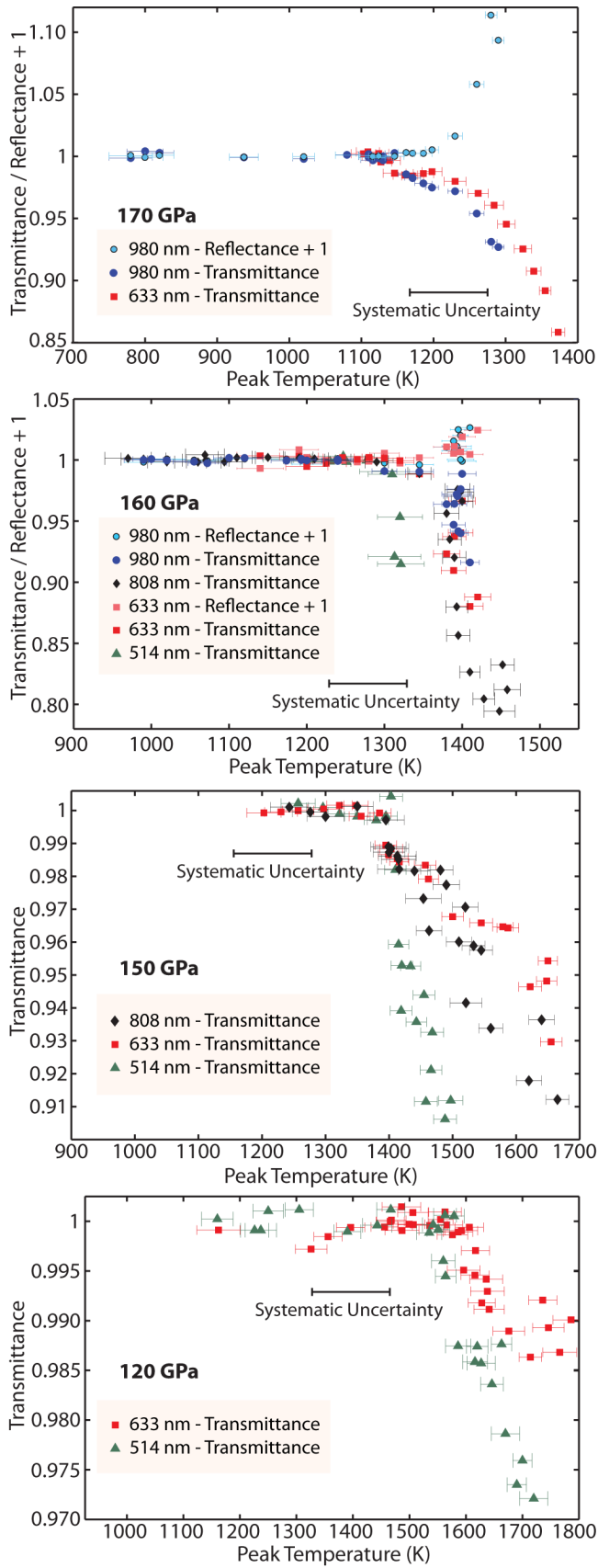

Fig. SI6. Collection of transmittance and reflectance data versus peak temperature at pressures. At a given plateau temperature corresponding to the plateaus in the heating curves, the data shows a drop in transmittance for the different wavelengths of the probe lasers. Notice that at the lower pressure points, the data shows less drop in transmission at increasing wavelength. The error bar on each point represents uncertainty in the blackbody fit, while the large error bar represents systematic error for the temperature values at the longer wavelengths.

Here we estimate the energy of dissociation. At 1 bar, the heat of dissociation, $E_{diss}$, of a hydrogen molecule is 4.67 eV. Upon compression, many-body interactions lower $E_{diss}$ of condensed systems and it is estimated to be ~3eV for dense hydrogen at 1.5 Mbar (*4*). At 150 GPa, the energy required to fully dissociate a monolayer of hydrogen molecules (20 micron diameter) compressed in our sample cavity is the number of molecules in a monolayer times the dissociation energy, $N \times E_{diss} = 1\times10^{-5}$ mJ, where N is calculated from the equation of state of hydrogen 0.34 g/cc, Ref. (*5*). The spacing between monolayers is 1.26 Å, so there are ~8 monolayers of MH in 1 nm, assuming complete dissociation. In our experiment, the energy per pulse of our heating laser ranges from 0.01-0.05 mJ. The energy required to raise the temperature of a monolayer of hydrogen molecules to ~1200 K at 150 GPa is small in comparison to that required to dissociate. Because we defocus our heating laser beam only part of the pulse energy is absorbed by the sample. Since we are able to adjust the pulsed laser power to generate a heating curve we can easily vary the peak temperature to resolve points below, at, and above the plateau.

One of the problems encountered with our measurements was the deterioration of the tungsten film with time when the pressure was above a megabar. As a consequence we studied 13 samples to ensure reproducibility, of which 12 gave consistent results. One could visually observe that the initially smooth W film <u>slowly</u> became textured within a few weeks, shown in Fig. SI7. Tungsten is known to form tungsten hydride at high pressure and temperature (*6*), which might be responsible. Other possibilities are stress in the underlying alumina layer that cause the texturing or due to the cupping of the diamond at high pressure. For some of our runs we covered the tungsten absorber with a layer of alumina and this effectively slowed the hydrogen diffusion. A general consequence of the deterioration was that pulsed laser power had to be increased to achieve the same measured temperature; if we considered the measurements to be unreliable the runs were terminated. The measurements at 130 GPa were conducted twice on two different samples, using tungsten+alumina coated absorbers. Each run yielded a different value for the plateau (Fig. SI7-c). Visual observation of the W absorber under a microscope showed serious deterioration of the film. The runs were terminated and data discarded.

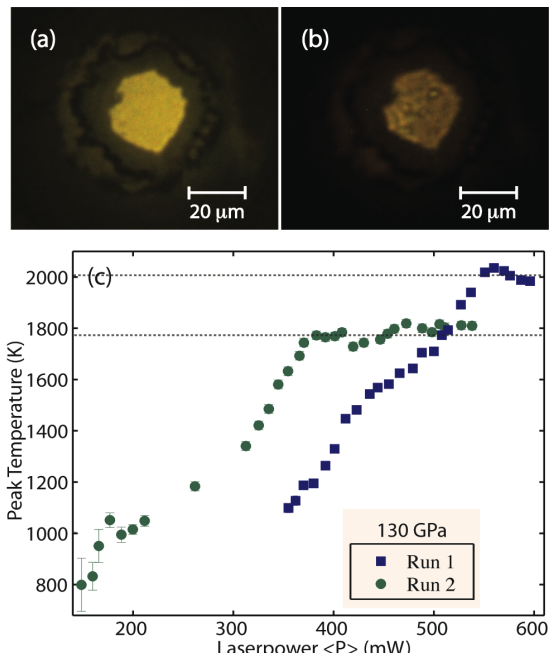

Fig. SI7. (a) The sample chamber loaded with hydrogen at 1Mbar for a typical run. One of the diamonds is covered with a semi-transparent W film deposited on a thin alumina layer. (b) The same sample chamber at ~1.3 Mbar and after many heating cycles. The W film exhibits deterioration/texturing. This could be due to chemical reactivity or the cupping of the diamond culets at these extreme conditions, which can strain the film. (c) Heating curves on deteriorated W films at the same pressure showing different values of the plateau temperature. Such data was discarded.